\documentclass[12pt]{article}
\def\cN{{\cal N}}

\newfont{\goth}{eufm10 scaled \magstep1}

\def\a{\alpha}

\def\c{\gamma}
\def\d{\delta}

\def\k{\kappa}
\def\l{\lambda}

\def\beq{\begin{equation}}\def\eeq{\end{equation}}
\def\beqa{\begin{eqnarray}}\def\eeqa{\end{eqnarray}}
\def\barr{\begin{array}}\def\earr{\end{array}}

\def \ys {{y\kern-.5em / \kern.3em}}
\def\ads{$AdS_5 \times S^5$ \,}
\def\R{{\bf R}}

\let\bm=\bibitem

\def\bd{\begin{document}}
\def\ed{\end{document}}
\def\ba{\begin{array}}
\def\ea{\end{array}}
\def\bea{\begin{eqnarray}}
\def\eea{\end{eqnarray}}
\def\ft#1#2{{\textstyle{{\scriptstyle #1}\over {\scriptstyle #2}}}}
\def\fft#1#2{{#1 \over #2}}
\newcommand{\be}{\begin{equation}}
\newcommand{\ee}{\end{equation}}
\newcommand{\eq}[1]{(\ref{#1})}
\def\eqs#1#2{(\ref{#1}-\ref{#2})}
\def\det{{\rm det\,}}
\def\tr{{\rm tr}}
\newcommand{\ho}[1]{$\, ^{#1}$}
\newcommand{\hoch}[1]{$\, ^{#1}$}
\def\ra{\rightarrow}
\def\uha{{\hat {\underline{\a}} }}
\def\uhc{{\hat {\underline{\c}} }}

\begin{document}

\hfill{NEIP-99-012}

\hfill{hep-th/9907106}

\vspace{20pt}

\begin{center}

{\Large\bf  A Note on the 
Chiral Anomaly in the AdS/CFT Correspondence and $1/N^2$
Correction }
\vspace{30pt}

{\large Adel Bilal and Chong-Sun Chu }

\vspace{15pt}

{\small \em Institute of Physics,
University of Neuch\^atel, CH-2000 Neuch\^atel, Switzerland}

\vskip .2in \sffamily{ Adel.Bilal@iph.unine.ch \\ cschu@sissa.it}
\vspace{60pt}

{\bf Abstract}
\end{center}

According to the AdS/CFT correspondence, 
the $d=4$, $\cN=4$ $SU(N)$ SYM is dual to 
the Type IIB string theory compactified
on \ads. A mechanism was proposed previously that the chiral anomaly
of the gauge theory is accounted for  
to the leading order in $N$
by the Chern-Simons action in the $AdS_5$ SUGRA. 
In this paper, we consider the SUGRA$\backslash$string 
action at one loop 
and determine  the quantum corrections to the Chern-Simons action.
While gluon loops do not modify the coefficient of the 
Chern-Simons action, spinor loops shift the coefficient 
by an integer. 
We find that for finite $N$, the quantum corrections from the complete
tower of Kaluza-Klein states reproduce exactly the desired shift 
$N^2 \rightarrow N^2- 1$
of the
Chern-Simons coefficient, suggesting that this coefficient does not 
receive corrections from the other states of the string theory.
We discuss why this is plausible. 

\pagebreak 

\section{Introduction }

According to the AdS/CFT correspondence \cite{c1,c2,c3,rev}, 
the $\cN=4$ $SU(N)$ supersymmetric gauge theory 
considered in the  `t Hooft limit with $\l \equiv g_{YM}^2 N$ fixed 
is dual
to the IIB string theory compactified on \ads. 
The parameters of the two  theories are 
identified as 
$g_{YM}^2 = g_s$, $\l = (R/l_s)^4$ and hence $1/N =g_s (l_s/R)^4$. 
In general, a gauge theory diagram with genus $g$ 
comes with a power of $N^{2-2g}$ and with some powers of
$\l$. In the string theory, correlation functions for a given genus
$g$ come with a power of  
$g_s^{2-2g}$ and have a coefficient given by a power series 
expansion in $\a'/R^2 =\l^{-1/2} $. 
At a given order of $N$ or $g_s$, 
the functions of $\l$ have different expansions (powers of $\l$ versus
powers of $\l^{-1/2}$) in the two theories,
although according to the AdS/CFT proposal,
they are supposed to represent the same function. 
It is clear that comparison between the two theories is
possible only for quantities whose $\l$ dependences 
can be computed exactly or that are independent of $\l$. 
Since the anomaly is independent of $\l$, 
it is a perfect candidate for this purpose. 

In the $\cN=4$ SYM, we have two kinds of anomaly, the trace anomaly 
and the $SU(4)_R$ chiral anomaly. Both have been checked 
\cite{c3,dzf,a1}  and found to
match with SUGRA calculations to leading order at large
$N$. Subleading orders in $N$ for other gauge systems have also been
discussed \cite{a2,a3}. However, the $1/N^2$ corrections for the
original $\cN=4$ case have never been tested.
In this paper, we will be interested in the $1/N^2$ corrections to the
chiral anomaly. To leading order in $N$, the chiral 
anomaly has been elegantly accounted for \cite{c3,dzf} 
by the Chern-Simons action in the $AdS_5$ SUGRA. The coefficient $k$
(see \eq{sugraaction} below) determines the magnitude of the chiral
anomaly in field theory and is given by $k=N^2$ to the leading order
in $N$. For finite $N$, the chiral anomaly of the $SU(N)$ gauge theory
is proportional to $N^2-1$ and there is a mismatch of -1. 
Our goal is to determine the $1/N^2$ correction to this coefficient
$k$ and reproduce this ``-1'' correction. 

According to the proposal
\cite{c1,c2,c3}, quantities of order $1/N^2$  corresponds to 
a 1-loop string calculation. Although a classical string action on
\ads 
with RR fields background is available \cite{adsstring}, 
the quantization of it is
notoriously difficult and we still don't have reliable means to
compute its  quantum corrections. This is partially reflected by the
fact that we still don't know the complete spectrum of states of the 
string theory on \ads. 
The only explicitly known states are the full towers of KK states 
\cite{vN2} coming from the compactification 
of the 10 dimensional IIB SUGRA multiplet. 
It is thus natural to first examine the quantum corrections to the
Chern-Simons coefficient coming from  all of
these states. In fact, we find precisely the expected correction of -1. 
This correction of -1 is entirely due to the quantum treatment of the
doubleton multiplet that is gauged away. This is very satisfactory
since the doubleton is known to be dual to the decoupled $U(1)$ factor
of the SYM theory which is $SU(N)$ rather than $U(N)$. 
Given that this already provide the desired shift $N^2 \rightarrow N^2
-1$, we expect that there is no contribution from  other
string states at all. We also note that 
a consistent truncation to the states of
the 5 dimensional supergravity multiplet alone is not sufficient. 

We will first review in sec. 2 the arguments  \cite{c2,dzf}
for matching the chiral anomaly with the Chern-Simons term in the five
dimensional SUGRA action to leading order in  $N$. Then in sec. 3 
we determine the one loop corrections  to the coefficient of the
Chern-Simons action from the 
Kaluza-Klein towers. We finally discuss why it is plausible that
other string states don't modify the Chern-Simons coefficient.

\section{Chiral Anomalies in the $\cN=4$ SYM}

The $\cN=4$ SYM in 4 dimensions has a R-symmetry group of
$SU(4)_R$. The matter fields are in nontrivial representation of it,
with the scalars $X^i$ transforming in the ${\bf 6}$ and the four complex
Weyl fermions $\l$ in  the fundamental representation of $SU(4)_R$
with the chirality part (0,1/2) in ${\bf 4}$ and (1/2,0) 
in ${\bf 4}^*$ (see for example \cite{FFZ}. 
Our convention here together with the convention we adopt 
in \eq{table} are equivalent to those these authors used.) 
It is convenient to use the compact  notation of differential forms 
\cite{zumino1,AGG}. 
The correctly normalized anomaly is derived from the descent equation
\be
\frac{i^{n+2}}{ (2\pi)^n (n+1)!} Tr F^{n+1} 
= d \omega_{2n+1} (A), \quad\quad
\d_v \omega_{2n+1} = d \omega^1_{2n} (v,A),
\ee
where 
\be
 F=dA+A^2, \quad\quad  \d_v A = dv + [A,v]
\ee
and
$ v=v^a T^a $, $A = A^a T^a$. $v^a$ are commuting gauge
parameters and $T^a$'s are anti-Hermitian and are in the fundamental
representation.
Explicitly for $n=2$, 
\bea
&& \omega^1_4 (v,A) = \frac{1}{24 \pi^2} Tr [v d(AdA +\frac{1}{2}A^3)],\\
&&\omega_5(A) = \frac{1}{24 \pi^2} Tr[ A(dA)^2 + \frac{3}{2} A^3 dA + 
\frac{3}{5} A^5 ]. 
\eea
In this notation,  the R-symmetry anomaly is 
\be \label{YManom}
v \cdot (D_i J^i) = - (N^2-1) \; \int_{S^4} \omega^1_4(v,A),
\ee
where we used the convention for Einstein summation:
\be
v \cdot F = \int d^{4}x \; v^a(x) F^a(x)
\ee
for any $v^a(x)$ and $F^a(x)$. 
The $N^2 -1$ factor is due to the 
fact that $\l$ is in the adjoint of $SU(N)$.

For $T^a$ in a general representation $\R$ of the group,
the corresponding quantities with the trace taken in $\R$ are
\be
{\omega^1_{2n}}^\R = A(\R) \omega^1_{2n}, \quad\quad
\omega_{2n+1}^\R =A(\R) \omega_{2n+1} ,
\ee
where $A(\R)$ is the anomaly coefficient defined by the ratio of the
$d$-symbol taken in the representation $\R$ and the fundamental
representation. In general $2n$ or $2n+1$ dimensions, since 
the $d$-symbol is given by a symmetrized trace of $n+1$ 
Lie algebra generators,  it is easy to show that the complex conjugate
representation ${\bf R}^*$ has an anomaly coefficient
\be \label{RR*}
A({\bf R}^*) = (-1)^{n+1} A({\bf R}). 
\ee

According to the AdS/CFT proposal, one should be able to see this
anomaly from the dual point of view of string theory. 
To  leading order in $N$, one looks at the IIB 
SUGRA compactified on \ads . The tree level SUGRA action contains the 
term \cite{GRW1,vN1,c3,dzf}
\be
\label{sugraaction}
S_{cl}[A]= \frac{1}{4 g^2_{SG}} \int d^5 x  \sqrt{g}\, 
 F_{\mu\nu}^a F^{\mu\nu a}  +
k \int_{AdS_5} \omega_5.
\ee
We note that in terms of components, 
$k \int \omega_5=\frac{i k}{96 \pi^2} \int d^5 x
d^{abc}\epsilon^{\mu\nu\lambda\rho\sigma}
A_\mu^a\partial_\nu A^b_\lambda\partial_\rho A^c_\sigma
+ \cdots $, 
if one uses  Hermitian generators and the definition 
of  $d$-symbol in \cite{dzf}.
From the dual gauge theory point of
view, since the current $J^a$ is coupled to the $SU(4)$ gauge fields
of the bulk,  using the AdS/CFT proposal one can determine 
the coefficient $g^2_{SG}$ and $k$
from the 2-point and 3-point correlators of $J^a$ \cite{dzf} of the
gauge theory.
The
normalization to leading order in $N$
\be \label{gkstd}
g_{SG}^2 = \frac{16 \pi^2}{N^2}, \quad k = N^2, 
\ee
has been determined in \cite{dzf}.
The ratio of the coefficient $g^2_{SG}$ and $k$  also agrees
with what one gets from supersymmetry \cite{GRW1,vN1}.
One may also determine  the values of $g_{SG}^2$ and $k$ 
from a dimensional reduction of the 10 dimensional 
IIB SUGRA. This requires the knowledge of the $N$ dependence of $R$. 
According to the
proposal \cite{c1,c2,c3}, the radius $R$ of $S^5$ is
determined by $R^4/\a'^2 = g_s^2 N$.  
Using this, it is easy to determine the 
normalization of the gauge kinetic energy term
and one indeed finds 
$g_{SG}^2 = {16 \pi^2}/N^2$ and hence
$k=N^2$ using  SUSY.

In usual consideration of SUGRA on $AdS$, one considers gauge
configurations $A^a$ which vanish at the boundary and so the 
Chern-Simons term is gauge invariant. For the consideration of AdS/CFT
correspondence, the boundary value of the $A^a$ is nonvanishing and 
coupled to the R-currents $J^a$. Under a gauge variation 
$\d_v A$, the variation of the 
Chern-Simons term is a boundary term 
\be \label{dS}
\d_v S_{cl} = k \int_{S^4} \omega^1_4.
\ee
Now by the conjecture~\cite{c1,c2,c3}, 
\be 
S_{cl}[A^a_{\mu}(x, x^5)]= \Gamma[A_i^a(x)], 
\ee
where $\Gamma$ is the generating functional 
for current correlators in the boundary theory, and hence 
\be
\d_v S_{cl} = \d_v  \Gamma = - v \cdot (D_i J^i) .
\ee
From this and \eq{dS}, one can read off the anomaly. It is
\be \label{GRanom}
v \cdot (D_i J^i) =- N^2  \; \int_{S^4} \omega^1_4 (v,A),
\ee
which agrees with the gauge theory computation \eq{YManom} to
leading order in $N$.

\section{Induced Chern-Simons}

As we have
seen in the previous section, the IIB  tree level SUGRA contains a
Chern-Simons term which can account for the chiral anomaly of the gauge
theory to leading order in $N$. But there is also a mismatch
of ``-1'' which is of order $1/N^2$.
In this section, we will examine the
$1/N$ corrections to the Chern-Simons action on the string theory side.
From the point of view of the IIB string theory,
an expansion in $1/N$ is a quantum expansion
beyond  tree level. In particular the $1/N^2$ correction to the
chiral anomaly corresponds to a  1-loop computation in IIB string
theory.
We will first examine the corrections coming from the
Kaluza-Klein states. Based on the origin of the Chern-Simons
action in $AdS_5$ supergravity, we will argue in the
discussion section that the other string states are not
likely to modify the Chern-Simons coefficient. 

\underline{Fermionic contributions}

It is well known that chiral fermions in even dimensions 
can give rise to an anomaly. Although there is no chirality in odd
dimensions, there is a similar phenomenon for fermions
in odd dimensions. Consider a Dirac fermion $\psi$ in odd 
dimensions (flat) minimally
coupled  to  (external or gauge) vector 
bosons $A_\mu$ of a group $G$. 
At the quantum level, a
regularization needs to be introduced to make sense of the theory and 
one cannot preserve both the gauge symmetry (small and large)  and the
parity at the same time \cite{fcs1,fcs2}.
If one chooses to preserve the gauge symmetry by doing a
Pauli-Villars regularization,  then there will be an induced
Chern-Simons term generated at one loop. The result is independent of
the fermion mass \cite{fcs2,AGPM}. 
In our notation, the
induced Chern-Simons term is 
\be \label{ind}
\Delta \Gamma = \pm \frac{1}{2}\int \omega^\R_{2n+1} = 
\pm \frac{1}{2} A(\R)  \int \omega_{2n+1},
\ee
where $\R$ is the representation of the Dirac 
fermion. The $\pm$ sign depends on
the regularization and can often be fixed within  
a specific context. 

This result was originally \cite{fcs1,fcs2} obtained  
for fermions coupled to gauge fields in a flat spacetime and has been 
extended to full generality for arbitrary curved backgrounds and any
odd dimensions 
\cite{AGPM}. The induced parity violating terms are
given (up to a normalization factor) 
by the secondary characteristic class 
$Q(A,\omega)$ satisfying
\be \label{Q}
dQ(A,\omega) = {\hat A(R)} ch(F) |_{2n+2},
\ee
where $\omega$ is the gravitational connection. 
Going one step down the descent relation, one gets 
the chiral anomaly of a Dirac operator
defined on a curved manifold.  
Since $ {\hat A(R)} = 1 + o(R^2)$
and $Tr F=0$ for $SU$, in five dimensions 
there is only the gauge Chern-Simons
and the gravitational Chern-Simons terms from \eq{Q} and there is no
mixing term.
It is clear that the pure gauge Chern-Simons piece takes the same
expression \eq{ind} as in the flat case. This can  indeed 
be expected from the
beginning as the gauge Chern-Simons form is independent of the metric.
It is also clear that there is no induced  gravitational Chern-Simons 
term as  it would be related to a  
gravitational anomaly in four dimensions, and gravitational anomalies
exist only in $4k+2$ dimensions.

Now we need the particle spectrum of the type IIB string theory on
\ads. The only explicitly known states are the KK states
coming from the compactification
of the 10 dimensional IIB SUGRA multiplet \cite{vN2}. So we will
examine them first. 
Particles in $AdS_5$ are classified by their unitary irreducible
representation of $ SO(2,4)$. Since $ SO(2,4)$ has the maximal compact
subgroup $SO(2)\times SU(2)\times SU(2) $, irreducible representations
are  labelled by the quantum numbers 
$(E_0, J_1, J_2) $. 
The complete  KK spectrum of the IIB SUGRA on \ads
was obtained in  \cite{vN2} together with information on the 
masses (in units of $1/R$)
\footnote{There seems to be a misprint for the masses of
the spin 3/2 in the table of \cite{vN2}.} 
and the  representation content under $SU(4)_R$. 
We reproduce these results in the following table 
where we also give the  
$SU(2)\times SU(2)$ content \cite{GRW2} for the fermionic towers.

\be \label{table}
\begin{array}{lclclr}
         & SU(2)\times SU(2) &\mbox{masses}& & SU(4)_R \\ 
\psi_\mu & (1,1/2) & k+3/2    &k\geq 0  &{\bf 4, 20, \cdots} 
&\leftarrow \\
	 & (1,1/2) &-(k+7/2)  &k\geq 0  &{\bf 4^*, 20^*, \cdots} \\ 
\l       & (1/2,0) &-(k-1/2)  &k \geq 1 & {\bf \;\;\;\, 20^*, \cdots}
&\leftarrow \\
         & (1/2,0) & k+11/2   &k\geq 0  &{\bf 4, 20, \cdots} \\
\l'      & (1/2,0) &-(k+3/2)  &k \geq 0 & {\bf 4^*,20^*,\cdots}
&\leftarrow \\
         & (1/2,0) & k+7/2    &k\geq 0  & {\bf 4, 20, \cdots}\\
\l''     & (1/2,0) &-(k+9/2)  &k\geq 0  & {\bf 36, 140, \cdots}\\
	 & (1/2,0) &k+5/2     &k\geq 0  & {\bf 36^*, 140^*, \cdots}\\
\end{array}
\ee

The fermions in this table are symplectic Majorana spinors. 
For simplicity, we have only listed half of the field content. 
The other half (``mirror'') consists of fields with  conjugate
$SU(4)_R$ content and with the 
$SU(2)\times SU(2)$ quantum numbers exchanged \cite{GRW2}. 
The first member of the  rows marked 
with an $\leftarrow$ together form the fermionic sector of 
the $\cN=8$ supergravity multiplet: 8 gravitini and 48 spin 1/2. 
We have chosen to list the ``left handed'' spinors here.
The ``chirality''  refers to the anti-de Sitter group $SU(2,2)$. 
Since  the 
Chern-Simons term in odd dimensions is related to the 
chiral anomaly in 
one lower dimension \cite{AGPM},
the ``chirality'' thus allows us to fix the $\pm$ sign of the induced
Chern-Simons term. In particular, the  ``right-handed'' 
(``left-handed'') spinor generates a Chern-Simons term 
with + (-)  sign in \eq{ind}.  We should also remember \eq{RR*} that
the conjugate representation ${\bf R}^*$ has an opposite anomaly
coefficient in five dimensions. 
To determine the net induced Chern-Simons
terms from all these states, we should also sum over 
the ``mirror''. 
Notice that  the ``right-handed'' sector
contributes exactly the same as the ``left-handed'' sector 
because relative to the ``left'' sector, they get one
minus sign from the ``chirality'' and another one 
from $A({\bf R}^*)$. Notice also that  
the induced Chern-Simons term from 
a symplectic Majorana spinor is half that of a Dirac spinor 
because it contains half as many degrees of freedom.
Therefore effectively we can just sum over
the spectrum in the ``left handed'' sector above using the formula
\be \label{ind1}
\Delta \Gamma =
- \frac{1}{2} A(\R)  \int_{AdS_5} \omega_{5}
\ee
as if they are Dirac spinors. 

It is clear that 
the towers of $\psi_\mu, \l', \l''$  do not generate a net
induced Chern-Simons term as the states in these towers
all comes in pairs with their $SU(4)$ content conjugate of each
other. 
However, as pointed out in \cite{vN2}, there is a missing
$k=0$ state (${\bf 4^*}$) in the first  tower of $\l$.  There are
similar ``missing states'' in the  bosonic towers. 
Together they are identified with the doubleton multiplet 
of $SU(2,2|4)$ which consists of a gauge potential, six scalars and
four complex spinors. These are nonpropagating  
modes in the bulk of $AdS_5$ and can be gauged away 
completely \cite{vN2,GNW}, which is the reason 
why they don't appear in the physical spectrum. 
These modes are  exactly dual 
to the $U(1)$ factor of the $U(N)$ SYM living on the boundary 
\cite{rev}. 
Since from the SYM point of view, 
the -1 correction to the chiral anomaly is due the
decoupling of this $U(1)$ to give $SU(N)$, we expect that from the
dual point of view of the $AdS_5$ string theory,  
the -1 correction would also find an explanation 
entirely in terms of  these gauge modes. 
Indeed, since there is an additional $k=0$ state 
(``left-handed'' and in ${\bf 4}$) in the tower of $\l$
which is not balanced out, there will be a 
Chern-Simons term  of -1/2 (in units of $\int \omega_5$)
coming from this state. This is only half of the desired result. 
However this is  not the complete story. 

Let us recall the doubleton multiplet is absent because it has been
{\it gauged away} \cite{vN2} by imposing the gravitino gauge fixing
condition (see also \cite{GNW} for the gauging in the 
case of $AdS_7 \times S^4$ case). 
Hence to properly quantize the system, one has to introduce the
corresponding Faddeev-Popov ghosts. These ghosts will give another
contribution of $-1/2$ to the Chern-Simons coefficient. Indeed the ghost
multiplet contains bosonic spinors in exactly the same $SU(4)_R$
representation as the spinors of the doubleton multiplet, i.e. in 
${\bf 4^*}$. Natively one would expect a contribution opposite to the
one of the unbalanced $k=0$ state (in ${\bf 4}$) of the $\l$ tower. 
However, since the ghosts have opposite statistics, they actually give
the same contribution, i.e. another -1/2. So altogether we get a total
induced Chern-Simons term of -1, 
\be
\Delta \Gamma =
-  \int_{AdS_5} \omega_{5},
\ee 
which is exactly the desired result. 
Notice that the induced
Chern-Simons (coming with a constant integer coefficient) 
is independent of the radius $R$ 
and this is consistent with the AdS/CFT proposal
since the anomaly and its corrections are independent of $\l$. 

\underline{Bosonic contributions}

There is another interesting effect related to the Chern-Simons action. 
It is known that in three dimensions, the gluons at one loop 
can modify the coefficient of the  Chern-Simons action by an integer 
shift. 
However, the
precise modification depends on the regularization scheme adopted 
\footnote{
We thanks R. Stora for a useful discussion about the issues of
regularization. 
}.
In five dimensions, it is easy to convince oneself that there is no 
induced Chern-Simons Lagrangian coming 
from the gluon loops.
The reason is simple. Suppose one adopt some regularization
scheme to compute the gluon loops that may generate an  
induced Chern-Simons term. Since the gluons are in the adjoint
representation of the gauge group,  the diagrams are  
proportional to the $d$-symbol of the adjoint representation, or $A({\bf
adj})$. For three dimensions, this is proportional to 
the quadratic Casimir, but it is zero in five dimensions. 
In general, \eq{RR*} says that $A({\bf R}^*) = A({\bf R}) $ for $4k+3$
dimensions and  $A({\bf R}^*) = - A({\bf R}) $ for $4k+1$ dimensions. 
Hence $A({\bf adj}) =0$ for the present case. That the gauge bosons do
not modify the Chern-Simons coefficient was already noted by Witten 
\cite{c2}. Moreover, since
the other members (${\bf 64, 175, \cdots}$) 
of the spin 1 towers \cite{vN2} are all in real
representations, they also don't modify the Chern-Simons action. 

We thus see that while gauge boson loops do not modify the magnitude
of the Chern-Simons action, the spinor loops do. Due to the intimate
relation with the chiral anomaly \cite{AGPM}, one expects that 
there is a nonrenormalization theorem (counterpart of the 
Bardeen nonrenormalization theorem for the chiral anomaly)
that protects the Chern-Simons action from 
further corrections beyond one loop. 

Combining the above, we find that at finite $N$, the
coefficient $k$ is shifted by
\be \label{kshift}
k \rightarrow k-1\quad\mbox{or}\quad N^2 \rightarrow N^2 -1
\ee
due to the quantum effects of the full set of Kaluza-Klein states.

\section{Discussion}

We have seen that the finite $N$ dependence of the $SU(4)_R$
chiral anomaly is exactly accounted for by the 
quantum effects of the full set of KK states. 
A  consistent truncation
to the 5 dimensional supergravity states does not do the job and has
no reason that it should. 
In fact, the total 
induced Chern-Simons term has its contribution coming  
from the $k=0$ state of the $\l$ tower and from the ghost spinors to
the doubleton multiplet, both have nothing to do with the five
dimensional 
supergravity multiplet.
Thus the -1 shift at finite $N$ of the chiral anomaly 
find a direct explanation entirely in terms
of the quantum effects of the doubleton multiplet in $AdS_5$,
which is known to be dual to the decoupled $U(1)$ factor of the gauge
theory. The whole picture is consistent. 
 
For the AdS/CFT
proposal to work at finite $N$, the other string states shouldn't
contribute any further corrections to the Chern-Simons coefficient.
While we don't have a proof of this statement, we would like to argue
that this is plausible. 

In \cite{harvey} the Kaluza-Klein origin of the
Chern-Simons action in gauged supergravity theory
was explored. In particular, it
was shown that for the case of  $AdS_7 \times S^4$ compactification of
the 11 dimensional SUGRA, the Chern-Simons term comes from the 
11-dimensional Chern-Simons coupling $\int C \wedge G \wedge G$ 
upon compactification. 
The origin of the Chern-Simons term in \ads SUGRA was also discussed and 
it is expected to arise similarly  
from the compactification of the 10 dimensional IIB SUGRA. 
However, there are some technical subtleties, the major one being
the lack of a covariant action of the IIB SUGRA. It would be very 
interesting to work it out explicitly, 
presumably the formalism of \cite{cov} will be useful. 

Given that the Chern-Simons term has its origin in the 10 dimensional
SUGRA multiplet, it is natural to expect that 
the quantum corrections to the Chern-Simons action are 
also  confined to the states related to the  Kaluza-Klein
compactification 
and we conjecture that this is indeed the case. 

Since the trace anomaly and the R-symmetry 
anomaly are in the same SUSY multiplet, a similar finite $N$ 
correction of -1 should also appear in the trace anomaly.
We expect that again the correction will have an explanation 
entirely in terms of the quantum effects related to 
the  doubleton multiplet. 
It would be interesting to verify this explicitly. 
This will provide a cross check of our conjecture here. 

Loop effects in $AdS_5$ supergravity are definitely highly
nontrivial to compute.
In fact,  tree level calculations already call for the
invention of many ingenuous techniques and methods.  
Were it not for the
topological character of the Chern-Simons action, 
one will not be able to determine exactly 
the parity violating term induced  at one loop.
For other processes involving loops, in addition to the
difficulty of evaluating  complicated momentum integrals, one also
has a lot  more states propagating since a truncation of states 
to the 5 dimensional supergravity multiplet is not enough. It would be
interesting to find other quantities that can be calculated at one or
more loops 
in order to provide other tests of the AdS/CFT proposal at finite $N$.  

Since Chern-Simons action exists generally  in gauged supergravity
on $AdS_{p+1}$ space, they all have interpretation as chiral
anomaly in the dual CFT. For example, in the $AdS_7\times S^4$ case
discussed in \cite{harvey}, the Chern-Simons action has a coefficient
of $N^3$ and the $1/N^2$ correction would be of order $N$. We
suspect that this correction would again find an explanation in terms
of the doubleton multiplet, which now consists of a tensor gauge field,
four spinors and five scalars.   

\bigskip
\vspace{.5cm}

\noindent{\large \bf Acknowledgments}

We would like to thank  L. Alvarez-Gaume, L. Bonora, J.-P. Derendinger, 
D. Matalliotakis, R. Russo, R. Stora and B. Zumino
for discussions. We are particularly grateful 
to  R. Russo for initial collaboration. Support by the Swiss National
Science Foundation is gratefully acknowledged.


\ed